\documentclass[12pt]{article}

\usepackage{epsfig}

\newcommand{\be}{\begin{equation}}
\newcommand{\ee}{\end{equation}}
\newcommand{\bea}{\begin{eqnarray}}
\newcommand{\eea}{\end{eqnarray}}

\newcommand{\BV}{\left(\begin{array}{c}}
\newcommand{\EV}{\end{array}\right)}
\newcommand{\BM}{\left(\begin{array}{cc}}

\newcommand{\NL}{\nonumber \\}                    

\begin{document}

\begin{titlepage}

\title{Occam's Higgs: A Phenomenological Solution 
to the Electroweak Hierarchy Problem}

\author{
T. Goldman\thanks{\small \em 
Mail Stop B283; E-mail: t.goldman@post.harvard.edu}~ and 
Michael Martin Nieto\thanks{\small \em 
Mail Stop B285; E-mail: mmn@lanl.gov}\\
{\small \em Theoretical Division, Los Alamos National Laboratory,} \\ 
{\small \em University of California, Los Alamos, New Mexico 87545 USA}}

\date{\today} 
\maketitle

\vspace{-4.4in}
\flushright{LA-UR-03-9171}
\vspace{-0.2in}
\flushright{hep-ph/0403027}
\vspace*{4.2in}

\begin{abstract}

We propose a phenomenological solution to the Electroweak 
hierarchy problem.  It  predicts no new particles beyond 
those in the Standard Model. The Higgs is arbitrarily 
massive and slow-roll inflation can be implemented naturally. 
Loop corrections will be negligible even for large cutoffs.  

\end{abstract}


\end{titlepage}


\pagebreak 
\setcounter{page}{2}
\parindent=.3in

The hierarchy problem~\cite{eldad} for the Electroweak 
component of the Standard Model (SM) has led to a number 
of theoretical conjectures towards its resolution, from 
Technicolor to Supersymmetry.  All have encountered 
increasing difficulties with experiments.  An alternative 
concept, deconstruction from a higher dimension, has been 
proposed~\cite{georgi,hill}.  Based on this idea, a 
phenomenological solution to the hierarchy problem has 
been put forward in Ref.\cite{cohen}.  This solution 
invents {\it only} 30 new degrees of freedom (particles 
and interactions) to solve the problem vs., for example, 
56 in Ref.\cite{pvs} and from 126 up to thousands in 
supersymmetric approaches. 

Why is it widely assumed that the mass of the Higgs 
necessarily implies the existence of (so) many additional 
degrees of freedom?  Such a direction of thought is 
motivated by theoretical prejudices which, although they 
are highly attractive and have been beneficial in 
advancing understanding in the past, may nonetheless 
perhaps not be applicable here.  As a result of this 
prejudice, even more modest, phenomenological approaches 
have not been fully explored.  We delve into this 
viewpoint here.

In particular, we ask what is the minimal (hence, ``Occam'')
implication of the (apparently large) mass of the Higgs 
compared to expectations in the SM?  We contend that the 
only things known with certitude are that the Higgs has a 
nonzero vacuum expectation value, $\langle vev \rangle$, 
and a possibly very large mass. We seek to represent this 
information phenomenologically.

We observe here that nothing at all new need be proposed, 
provided that one is willing to describe the Higgs sector as 
an effective theory\footnote{For a general analysis in terms 
of renormalizable theories where loop corrections must be 
implemented, see Ref.\cite{AB}.  Much additional work was 
stimulated by this paper.} with unknown (and, we suspect, 
unknowable without much new data) underlying degrees of 
freedom.  That is, we accept a non-renormalizable Higgs 
potential.\footnote{This attitude is similar in spirit to 
the ``super-weak'' line of thought developed in connection 
with CP-violation by Wolfensten \cite{LW}.}  

We take the phenomenological constraints to be two-fold: 
(1) the $\langle vev \rangle$ of the Higgs should be stable, 
and (2) the mass of the Higgs should be arbitrarily large. 
These objectives can be straightforwardly achieved through 
nonlinearity and the abandonment of renormalizability as a 
constraint. We would be tempted to call what we present here an 
"effective field theory" but that terminilogy is already 
spoken for~\cite{Weinberg}.  It has specific meanings in 
terms of relations to underlying physics and degrees of 
freedom. 

What we have in mind may be thought of as a possible (path integral) 
solution of an underlying field theory represented in terms of 
effective degrees of freedom. A source for the (composite) 
Higgs is introduced and path integration carried out to produce 
all of the n-Higgs vertex functions. These include all quantum 
corrections and fully describe all self-interactions of Higgs 
bosons. (We ignore vertex functions involving coupling to 
other observed degrees of freedom.) As such, the Lagrangians 
described below should have complex momentum dependences in 
the coefficients of the n-Higgs vertex functions. We approximate 
these crudely by simple constants. 

An objection might be raised regarding the question of the 
constraint relating renormalization, mass, and unitarity. 
However, as has been observed before (see e.g., Ref.\cite{CLT}), 
this relation is based on satisfying unitarity perturbatively.
In fact, without self-shielding effects, unitarity is violated 
at a pole location unless the width of the particle involved is 
sufficiently large.  Using an N/D method, Lee et. al.~\cite{bwl1,bwl2} 
enforced unitarity without a perturbative requirement. They found 
that the effective Higgs' mass and width both increase significantly, 
but did not encounter any consistency problem. The effective 
interaction becomes strong and shelf-shielding to preserve the 
unitarity bound that MUST be satisfied. This phenomenon is familiar 
from the theory of Regge trajectory exchange~\cite{regge} applied 
to strong interaction physics. 

The $\langle vev \rangle$ may be fixed by replacing the SM 
Higgs by a nonlinear representation, analogous to that in 
the nonlinear sigma model. In the $O(n+1)$ sigma model, one 
replaces the invariant
\be
{\cal L}_0 = 
(\sigma - {\sigma}_{0})^{2} + \sum_{i=1}^{n} {\pi_{i}}^{2}, 
\ee 
where ${\sigma}_{0}$ is the sigma field  $\langle vev \rangle$, 
by its nonlinear equivalent~\cite{AD}
\be
{\cal L}_1 = 
\frac{\sum_{i=1}^{n}{\pi_{i}}^{2}}
{\sqrt{({\sigma}_{0})^{2} + \sum_{i=1}^{n}{\pi_{i}}^{2}}},
\ee
thereby disallowing any fluctuations of the $\sigma$ field by 
fiat. The cost is the appearance of a nonrenormalizable theory, 
which is then viewed as an effective field theory approximation 
to some underlying one. Formally, this is described by a group 
ratio, $SO(n+1)/SO(n)$ in this example. 

\begin{figure}[h!]
 \begin{center}
\noindent    
\psfig{figure=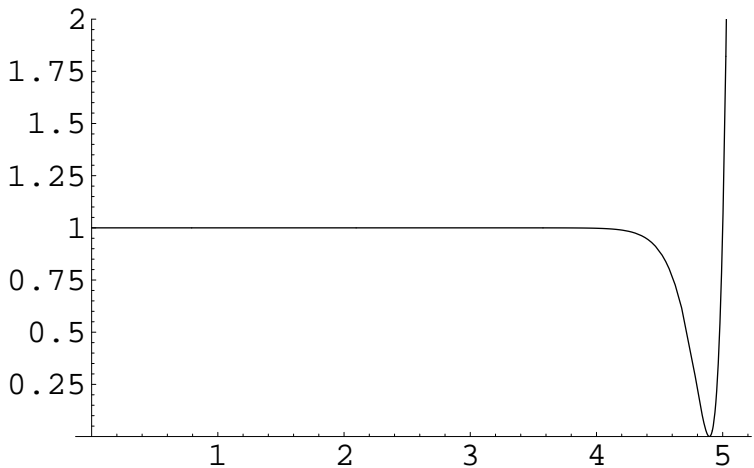,width=4in}
\end{center}
\caption{ $V(x)/\Lambda^4$ vs. $x$ for  $(a=10,~b=9,~c=5)$.
\label{pot}}
\end{figure}

\begin{figure}[h!]
 \begin{center}
\noindent    
\psfig{figure=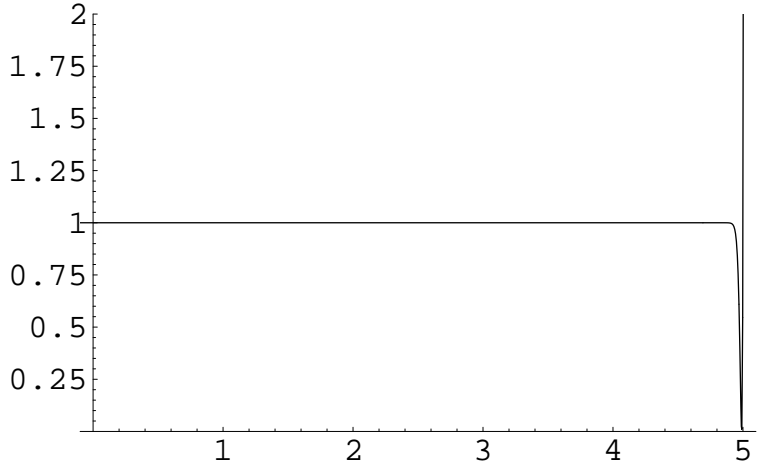,width=4in}
\end{center}
\caption{ $V(x)/\Lambda^4$ vs. $x$ for  $(a=100,~b=99,~c=5)$.
\label{potb}}
\end{figure}

A slightly less drastic approach allows for fluctuations of the 
$\sigma$ field, again at the cost of introducing nonrenormalizable 
interactions. In particular, let the Lagrangian density 
\be
{\cal L}_2 = 
\{ \sigma^{2} + \sum_{i=1}^{n} {\pi_{i}}^{2} - 
	({\sigma}_{0})^{2}\}^{2}
\ee 
be replaced by 
\bea
{\cal L}_3(x) = V(x) &=& {\Lambda}^{4} \left\{ 1 + 
N\left[e^{a(x-c)}-e^{b(x-c)}\right], \right\} ~~~a>b,\\
N &=& \left( \frac{b}{a-b} \right) \left( \frac{a}{b} 
                 \right)^{a/(a-b)},\\
x &\equiv& \sqrt{(\sigma^{2} + \sum_{i=1}^{n} {\pi_{i}}^{2})}, 
\eea
where ${\Lambda}$ sets the arbitrary mass scale.  (Note, 
therefore, that $(\Lambda, ~ x, ~c, ~1/a, ~1/b)$ all 
have units of mass.) In Figs.\ref{pot} and \ref{potb} 
we show $V(x)/{\Lambda}^{4}$ 
vs. $x$ for $(a=10, ~b=9, ~c=5)$ and  $(a=100, ~b=99, ~c=5)$,
respectively (where the values of $a$ and $b$ are defined in 
units of ${\Lambda}^{-1}$ and those of $c$ are in terms of 
${\Lambda}$).



With only four parameters, $(\Lambda,~ a,~b,~c)$, $V(x)$ 
can fix the value of $\langle vev \rangle$ completely, 
independently of the curvature about the minimum in the 
field potential. As the latter determines the mass of the 
field representing fluctuations about the $\langle vev \rangle$, 
that mass may now be chosen freely to match the results of 
experiments. 

Specifically, the $\langle vev \rangle$ is given by 
the value of $x_0$ where 
\be
0 = \left( \frac{\partial V}{\partial \sigma}\right)_ {x=x_{0}}
= V^{\prime}(x_0)
\left(\frac{\partial x}{\partial \sigma}\right)_{x=x_{0}}.
\ee
Since at $x_0$, with $\pi_{i} = 0$, 
\be
\left(\frac{\partial x}{\partial \sigma}\right)_{x=x_{0}} = 1, 
\ee
one has
\be
\langle vev \rangle = x_0 = c - \frac{{\rm ln}(a/b)}{(a-b)}. 
\ee
Note also that
\be
\lim_{(a-b) \rightarrow \delta \ll (a,b)} x_0 
 = \left(c -\frac{1}{b}\right)~~
  {}_{\stackrel{{\textstyle \rightarrow}}{b \rightarrow \infty}} ~~ c.
\ee

The mass-squared is then given by 
\bea
m^2 & = & \left(\frac{\partial^2 V}{\partial^2\sigma}\right)_{x=x_0} 
= V^{\prime\prime}(x_0) \NL
& = & {\Lambda}^{4} 
      N \left[a^{2}e^{a(x_0-c)} - b^{2}e^{b(x_0-c)}\right] \NL
& = & ab~{\Lambda}^{4}, \label{mass}
\eea
since $V^{\prime}(x_o)=0$. 

For any of $\Lambda$, $a$, and $b$ separately arbitrarily 
large, the value of the mass can be pushed to arbitrarily 
large values, overwhelming quadratically divergent corrections 
from additional interactions whatever the scale of the 
cutoff. From our point of view this is irrelevant anyway, 
as the theory is the approximate solution to the full result 
of path integration over the (unknown) underlying degrees of 
freedom. 

It is straightforward to carry out this procedure for the 
weak isoscalar Higgs isodoublet field of the SM. The result 
continues to suggest that underlying degrees of freedom exist, 
but abandons the notion that those of the SM necessarily 
represent fundamental quantities. Implicitly this applies to 
the quarks, leptons, and gauge bosons as well, as they may or 
may not themselves be fundamental degrees of freedom. Since 
string theory and even supersymmetry allow for this possibility, 
the suggestion made here should not be considered radical. 

Interestingly, potentials similar in form to that suggested 
here appear in some approaches to vacuum stabilization in 
string theory.~\cite{yuri,F} Further, potentials based on 
exponential forms have often been suggested within the 
context of the cosmological constant problem and 
quintessence~\cite{sahni}.

A side benefit of this kind of potential is that it is of the 
type that is required by studies of inflation~\cite{guth} as a 
solution to the flatness problem of the Universe. This occurs 
because 
\be
V^{\prime}(0) = {\Lambda}^{4} 
                N \left[ a e^{-ac} - b e^{-bc} \right]
\ee
behaves, in the limit $b \rightarrow a$, as 
\be
V^{\prime}(0)
           ~ \rightarrow~  -a(ac-1)~e^{-(ac-1)}~ {\Lambda}^{4},
\ee
which is arbitrarily exponentially small as $c \rightarrow \infty$. 
Note that this does not impinge upon the mass-squared value in
Eq. (\ref{mass}).

In conclusion, we ask what 
if the electroweak loop corrections 
were not already included in the conjectured approximate 
solution that we present for the path integral of whatever 
the underlying theory may be?   We answer that, 
with this potential and for sufficiently large $m^2$, 
electroweak loop corrections could be 
maintained as negligible even for cutoffs much larger than 
the $1~-~10$~TeV generally considered at present.

We thank Alex Friedland, Emil Mottola, Jos\'e R. Pel\'aez, 
John Terning and Yuri Shirman for helpful comments. 
This research was supported by the US Department of 
Energy under contract W-7405-ENG-36.


\end{document}